\documentclass[10pt,twocolumn,letterpaper]{article}

\usepackage{wifs}
\usepackage{times}
\usepackage{epsfig}
\usepackage{graphicx}
\usepackage{amsmath}
\usepackage{amssymb}

\usepackage{amsmath}
\usepackage{amssymb}
\usepackage{graphicx}
\usepackage{mathtools}

\usepackage{mathtools}
\usepackage[ruled,linesnumbered,vlined]{algorithm2e}
\usepackage{subcaption}
\usepackage{fancyhdr}

\usepackage[pagebackref=true,breaklinks=true,letterpaper=true,colorlinks,bookmarks=false]{hyperref}

 \wifsfinalcopy 

\def\wifsPaperID{92} 

\ifwifsfinal\fi
\begin{document}

\title{A Randomized Kernel-Based Secret Image Sharing Scheme}

\author{Akella Ravi Tej ~~~  Rekula Raviteja ~~~ Vinod Pankajakshan\\
Indian Institute of Technology Roorkee, 
Uttarakhand, India\\
{\tt\small ravitej.akella@gmail.com~~~}
{\tt\small rekula3@gmail.com~~~}
{\tt\small vinodfec@iitr.ac.in}
}

\maketitle

\begin{abstract}
   This paper proposes a ($k,n$)-threshold secret
image sharing scheme that offers flexibility in terms of meeting
contrasting demands such as information security and storage
efficiency with the help of a randomized kernel (binary matrix)
operation. A secret image is split into $n$ shares such that any
$k$ or more shares ($k\leq n$) can be used to reconstruct the image. Each share has a size less than or at most equal to the size of the secret image. Security and share sizes are solely determined by the kernel of the scheme. The kernel operation is optimized in terms of the security and computational requirements. The
storage overhead of the kernel can further be made independent
of its size by efficiently storing it as a sparse matrix. Moreover, the scheme is free from any kind of single point of failure (SPOF).
\end{abstract}

\thispagestyle{fancy}

\fancyhf{}

\renewcommand{\headrulewidth}{0pt}

\chead{\small Accepted for publication in IEEE International Workshop on Information Forensics and Security (WIFS) 2018, Hong Kong}

\pagestyle{empty}

\lfoot{\small \copyright 2018 IEEE. Personal use of this material is permitted. Permission from IEEE must be obtained for all other uses, in any current or future media, including reprinting/republishing this material for advertising or promotional purposes, creating new collective works, for resale or redistribution to servers or lists, or reuse of any copyrighted component of this work in other works.}

\section{Introduction}
In storage systems, information security is a top priority
alongside efficient storage. Most of the methods that have
been proposed with enhanced security have single storage
mechanism, which makes the entire system vulnerable if the
stored data is corrupted. This makes the storage system a
single point of failure (SPOF) for the entire scheme. To avoid
this vulnerability, ($k,n$)-threshold secret sharing techniques
are used to divide the information across $n$ participants such
that the shares of $k$ or more participants can reconstruct the
secret. Secret sharing has been an active area of research in the field
of efficient and secure storage of data since late 20th century
and has seen a lot of advancements in the recent decades.

Some of the first threshold secret sharing schemes
proposed were Blakley \cite{Blakley79} which involves a geometric approach and Shamir \cite{Shamir79} which uses polynomial interpolation. Later Mignotte \cite{Mignotte83} and Asmuth-Bloom \cite{Bloom83} came up with similar threshold schemes based on Chinese remainder theorem. Ito \textit{et al.} \cite{Ito87} proposed a more general secret sharing scheme with access structures to denote the subsets of all participants who can form a qualified group for secret reconstruction.

A ($k$,$n$)-threshold secret sharing scheme which leaks no information of the secret to an unqualified group of $k-1$ or fewer participants is called \textit{Perfect Secret Sharing} (PSS), suggested by Karnin \textit{et al.} \cite{Karnin83}. When exposed, information of the secret is tantamount to the amount of unqualified coalition of secret shares, the type of scheme is called \textit{Ramp Secret Sharing} (RSS) \cite{Okada94}. The classification between PSS and RSS is suggested in \cite{Bai06-1}. While the methods proposed in \cite{Blakley79} and \cite{Shamir79} are PSS schemes, \cite{Bloom83} does not ensure perfect secrecy of the shares. Bai \cite{Bai06-2} proposed an RSS scheme that performs matrix projection using randomly generated matrices.

Thien and Lin \cite{Lin02} proposed how Shamir's ($k,n$)-threshold scheme can be extended to images by reducing the size of secret shares by a factor of $k$ relative to the original image size. Though this method is very efficient in reducing the size of shares, it is not lossless due to truncation distortion. Moreover, the method is not a PSS scheme because there is a possibility of interpreting the secret information from compromise of $k-1$ or fewer shares. This drawback arises because of high spatial correlation of pixels in natural images. Though permuting image pixels as pre-processing can reduce this correlation, it cannot be considered as a fool-proof solution to the problem. Kuang-Shyr Wu \cite{Wu13} proposed an elegant method to overcome the truncation distortion in Thien-Lin's approach without using extra memory. This method exploits the high spatial correlation of pixels in natural images to its advantage.

Group signature, group encryption and secure multi-party computation employ a variant of secret sharing that also performs user authentication, known as Verifiable Secret Sharing (VSS), proposed by Chor \textit{et al.} \cite{Chor85}. To prevent the malicious modification of shares, there are methods proposed for cheating prevention \cite{Yan15} and verifiability \cite{Lin10,Harn14}. Visual Cryptography Schemes (VCS), introduced by Naor and Shamir \cite{Naor94} is another active research area that uses secret sharing. Multiple-Secret VCS \cite{Shyu13}, XOR based VCS \cite{Yang14}, General access structure in VCS \cite{Shyu15}, Region Incrementing VCS and Fully Incrementing VCS \cite{Chen17} are some variants of VCS.

Despite the shortcomings, most of the present-day secret image sharing schemes use Thien-Lin's \cite{Lin02} method because of the high efficiency offered in terms of share size. While simplicity and storage efficiency are very desirable features, patching up major vulnerabilities of this method has motivated our research. The proposed method offers some improvisations to obviate the drawbacks that arise due to spatial correlation with a controlled increase in the share sizes.

Rest of the paper is organized as follows: ($k,n$)-threshold secret sharing mechanisms are discussed in Section 2. The proposed method is illustrated in Section 3. An adversarial model for the scheme is described in Section 4. The computational complexity and security analysis for the proposed method is done in Section 5 and Section 6. Finally the conclusions are summarized in Section 7.

\section{Preliminaries -- Secret Sharing Schemes}
Secret sharing is a way to distribute secret information across multiple participants such that only a qualified subset of participants can reconstruct the secret by pooling their shares. Although many approaches have been proposed to achieve ($k,n$)-threshold secret sharing schemes, polynomial schemes remain the most widely used. These schemes sample points from a polynomial for share generation and use Lagrange's interpolation for reconstruction. 

\subsection{Shamir's ($k,n$) Secret Sharing Scheme \cite{Shamir79}}
Shamir's ($k,n$)-threshold secret sharing technique uses a ($k -1$)$^{\text{th}}$ degree polynomial function defined as
\begin{equation}\label{eq1}
f(x) = (d_0 + d_1 x + d_2 x^2 +\cdots+ d_{k-1} x^{k-1}) \text{ mod} \ p
\end{equation}
where  $d_0$ is the secret information, $p$ is a prime number, and $d_1,  d_2,  ...,d_{k-1}$  are random numbers.
Let  $y_i = f(x_i),  1 \leq i \leq n$, and $0 <x_1 <x_2 \cdots <x_n<p$ , then the pairs ($x_i,y_i$) form the secret shares. Shamir's scheme is based on the fact that atleast $k$ distinct points on a ($k -1$)$^{\text{th}}$ degree polynomial are required to reconstruct the polynomial, making it a PSS. Alternatively, with $k$ shares, it is possible to construct $k$ linear equations in $k$ variables which have a consistent solution. Lagrange's interpolation is used to find the coefficients of $f(x)$ and subsequently the secret information. This scheme can be directly extended to images by taking each pixel at a time as secret information. As each pixel has a separate polynomial equation, each of the $n$ generated shares have the size of original image, making the scheme storage inefficient. It has to be noted that different random numbers are generated for each pixel's share equation.

\subsection{Thien-Lin's Secret Image Sharing Scheme \cite{Lin02}}
Thien-Lin's method proposes a memory efficient way to extend Shamir's \cite{Shamir79} method to images. Instead of taking one pixel at a time, the method suggests substituting all the $k$ coefficients with pixel values in \eqref{eq1}. Taking $k$ elements at a time to generate $n$ shares, makes the total size of shares $n/k$ times of the original image, with each share $1/k$ times the size of original image. Image pixels are chosen sequentially or according to a permutation cipher, as the coefficients of the polynomial in \eqref{eq1}. Once the share images are generated, the polynomial is destroyed and the share images are distributed among the participants. During reconstruction, with $k$ or more participants, the generated pixel intensities must be sequentially mapped to the secret image in the order in which they were selected.

Although the method is storage efficient, it has some drawbacks. The reconstruction is not lossless as all the pixel values above 250 are truncated to 250 (The value of $p$ in \eqref{eq1} is taken as 251, the largest prime number less than 256). This is called \textit{truncation distortion}. Also, Thien-Lin's approach is not a PSS as less than $k$ shares can provide partial information about the secret image. This arises because of high spatial correlation in natural images. Though one cannot exactly determine all the $k$ pixel values, it is possible to estimate pixel intensities as all the coefficients of the polynomial are approximately in the same range. Hence, the polynomial in \eqref{eq1} can be modified as 
\begin{equation}\label{eq12}
f(x) \approx \big(d_0(1 + x + x^2 +\cdots+ x^{k-1})\big) \text{ mod} \ p
\enspace.
\end{equation}

\noindent
Given one share pair ($x_i,y_i$), we can guess $d_0$ which is nearly same as $d_1,d_2,…,d_{k-1}$ due to high spatial correlation. Note that more shares would lead to more accurate predictions and hence fail the secrecy of the scheme.

\subsection{Kuang-Shyr Wu Secret Image Sharing Scheme \cite{Wu13}}
Kuang-Shyr Wu suggested a modification to Thien-Lin's \cite{Lin02} method to avoid the truncation distortion.  Wu's method suggests the value of $p$ to be chosen as 257 ($p$ is taken as 251 in Thien-Lin's method). The share values generated lie in the interval of 0 and 256 (requires 9 bit encoding), but the image intensities lie in the interval of 0 and 255 (8 bit encoding). The need for additional memory is avoided by treating 256 as a special case and storing it as a 0. These shares are then distributed across participants and the polynomial is destroyed. During reconstruction, whenever a 0 is encountered, different combinations of 0 and 256 are used to generate candidate solutions. The best candidate solution is the one with the least total squared-euclidean distance of every pixel to its neighbours.

\section{Proposed Method}
Thien-Lin's \cite{Lin02} method with Wu's \cite{Wu13} modification would not be considered as a PSS as the correlation between pixels could leak some information of the secret image. Although using a pseudorandom permutation cipher as suggested in \cite{Lin02} reduces the spatial correlation between coefficients of the polynomial, the cipher becomes an SPOF for the entire scheme \cite{Bai06-1}. Furthermore, implementation of a pseudorandom permutation cipher as preprocessing in Thien-Lin's method is specific to a particular image dimensions and requires memory of the order of image size. Additionally, this process always generates the same permutation of pixels and subsequently the same shares for a given image. The system can be made less vulnerable by making some of the coefficients of the polynomial random, instead of image pixel values. Such a method however will be effective only if the number of random coefficients and their positions in the polynomial remain variable and unknown to the attacker. However, the randomness introduced must be perfectly predictable using a key for reliable reconstruction. In the proposed method, a kernel (binary matrix) is chosen as the key, through which one can accordingly vary the amount of randomness to be introduced which scales the security and share size. Details of the proposed method are described in the following subsections.

\begin{figure}[!ht]
    \centering
    \includegraphics[width=8.5cm]{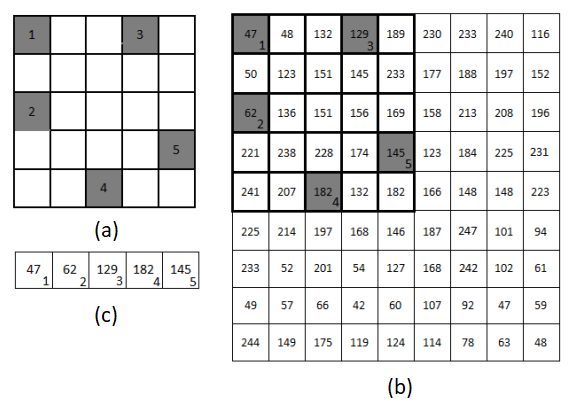}
    \caption{(a) Example of a $5\times5$ kernel, $k$=5, with $C = 1.2$ and gray cells corresponding to ones in the kernel; (b) $9\times9$ image matrix with cells values corresponding to intensities, the kernel is placed in its initial position; (c) $k$ values picked from the corresponding kernel position in (b).}
    \label{fig1}
\end{figure}

\subsection{Kernel Properties}
The kernel must follow certain properties for efficient share generation. It must have a total of $k$ ones and rest all cells are zeros, where $k$ is the threshold of the scheme. The first element in the kernel is one and the rest $k-1$ ones are randomly distributed in the kernel as shown in Fig. \ref{fig1}(a). A kernel with large dimensions for a given $k$ value provides greater security with the same memory and computational requirements (by storing the kernel as a binary sparse matrix).

\subsection{Kernel Operation}
The kernel operation shifts the chosen kernel over the image ensuring that each pixel is selected only once. At each position of the kernel, $k$ elements are selected corresponding to the $k$ ones in the kernel, i.e. pixel intensities that correspond to the ones in the kernel are picked, as shown in Fig. \ref{fig1}(b) and Fig. \ref{fig1}(c). Picking the same pixel twice is avoided by replacing it with a random number. Similarly, when the kernel goes partially outside the image bounds, random numbers are assigned for the ones that do not correspond to any image pixel. Disclosure of pixel data to attacks that leverage the high correlation in images is averted as the random numbers plummet the correlation between coefficients in the polynomial. This advantage in terms of security comes with a marginal increase in share sizes as the shares now hold the data of the randomly generated numbers in addition to pixel data. With controlled increase in share size, the kernel operation will prevent loss of pixel information through attacks that rely on the high spatial correlation of image pixels.

\begin{figure}[!ht]
    \centering
    \includegraphics[width=6.5cm]{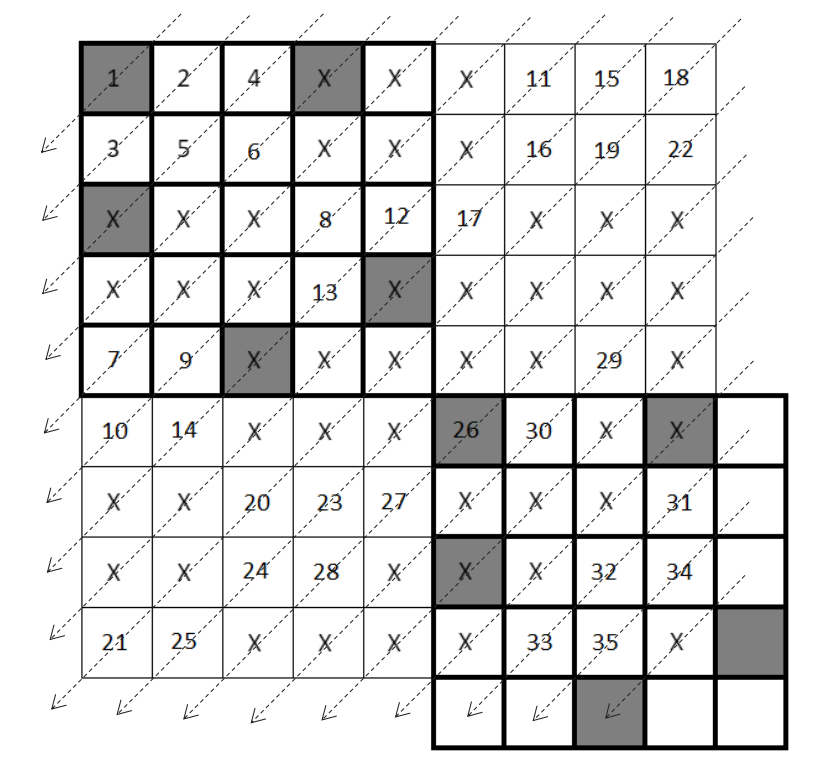}
    \caption{A $9\times9$ block same as the dimensions of the image in Fig. \ref{fig1}(b), showing the sequence of positions taken by the kernel in the order of numbering. X denotes previously used pixels which cannot be part of the sequence of positions taken by the kernel. The dotted arrows are used to show the order in which one must search for the nearest unmarked pixel (least Manhattan distance). The first and the $26^{th}$ positions of the kernel have been shown. Note that the $26^{th}$ kernel position has two out-of-image-bounds pixel selections.}
    \label{fig2}
\end{figure}

\begin{figure*}[!h]
\centering
\includegraphics[width=15cm,height = 3.75cm]{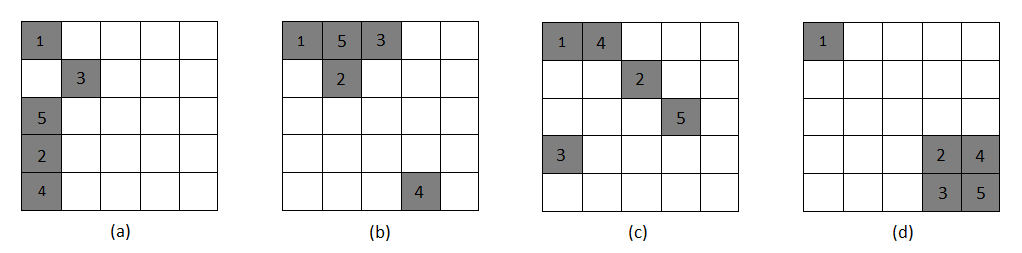}
\caption{Example $5\times5$ kernels with $k = 5$. Gray pixels represent the ones in the kernel with numbering numbering denoting the order in which pixels are selected. (a) $C$ = 0.2; (b) $C$ = 0.8; (c) $C$ = 1.0; (d) $C$ = 3.2.}
\label{25}
\end{figure*}

To prevent further storage overhead and for reliable reconstruction, the sequence of positions taken by the kernel must be derived from the kernel itself. One efficient approach to generate the sequence of positions (ensures minimum share sizes for a given kernel) is to initially place the kernel such that the first pixel of the image and that of the kernel coincide. At each position, all the image pixels that coincide with the ones in the kernel are marked. Now the kernel is shifted to a new position such that the first pixel of the kernel coincides with the nearest pixel to the first pixel (least Manhattan distance) of the image that is not marked previously. Subsequently the kernel is moved to positions farther from the first pixel sequentially until all the pixels in the image are covered. The sequence of positions taken by the kernel in Fig. \ref{fig1}(a) over the image Fig. \ref{fig1}(b) is shown in Fig.\ref{fig2}.

\begin{figure}[!ht]
\begin{subfigure}{0.5\textwidth}
\includegraphics[width= 8.75cm]{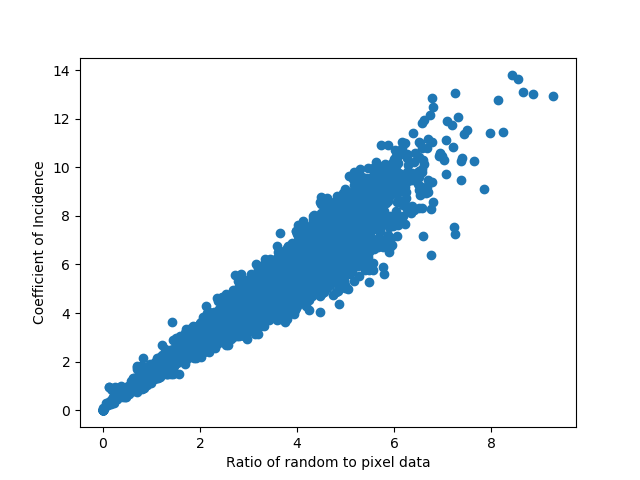} 
\caption{Plot between $C$ and the ratio of random to pixel data for multiple $45\times45$ kernel with varying $k$ values over images of various sizes (correlation coefficient = 0.9407).}
\label{fig45}
\end{subfigure}
\begin{subfigure}{0.5\textwidth}
\includegraphics[width=8.75cm]{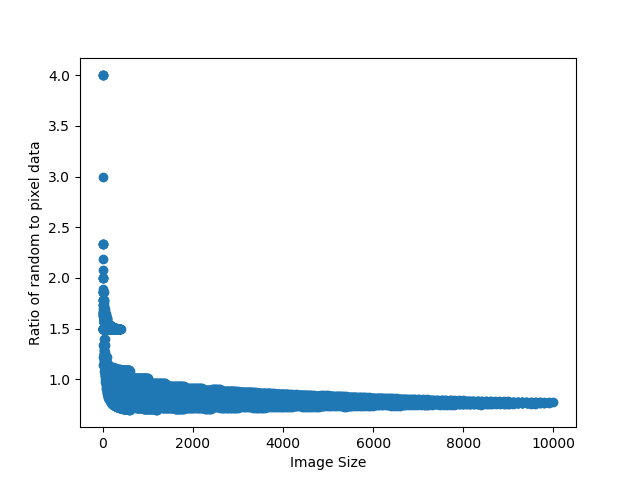}
\caption{Plot between the ratio of random to pixel data and image size for the kernel in Fig. \ref{25}(d). Here image size refers the number of pixels present in the image.}
\label{fig9}
\end{subfigure}
 
\caption{Scatter plots denoting the relationship between $C$, the ratio of random to pixel data and image size.}
\label{fig945}
\end{figure}

\subsection{Coefficient of Incidence}
We define \textit{coefficient of incidence} ($C$)  for a kernel as the ratio of number of random numbers generated to pixel data, when the kernel operation is performed on an image having the same dimensions as the kernel. We describe an iterative approach \textbf{Algorithm~\ref{Algo}} to estimate the $C$. The value of $C$ for a kernel can vary from $0$ (no random coefficients, scheme reduces to Thien-Lin's \cite{Lin02} method) to $k-1$ (theoretical limit, scheme reduces to Shamir's \cite{Shamir79} method for sharing images). Example $5\times5$ kernels and their corresponding $C$ values are shown in Fig. \ref{25}. It has to be noted that although the $C$ of Shamir's scheme is $k-1$ (as for every pixel, we select $k-1$ random numbers to generate the share polynomial), there does not exist a kernel that satisfies the constraints. As shown in Fig. \ref{fig945}, the number of random coefficients in the polynomial relative to the number of image pixels selected is highly correlated to $C$ (correlation coefficient = 0.9407 in Fig. \ref{fig945}(a)) and broadly independent of image size (with increase in image size, the variability in the ratio of random to pixel data decreases). This empirically shows that $C$ is a good metric to compare kernels based on the amount of randomness they produce when operated over an image of unknown dimensions. Greater the value of $C$, greater the number of random coefficients in the polynomials for a given image size, leading to greater share sizes.

\begin{algorithm}[!h]
\SetAlgoLined
 \caption{Coefficient of Incidence}\label{Algo}
  \SetKwInOut{Input}{input}
  \SetKwInOut{Output}{output}

  \Input{$X$ is the set of positions of all ones in the kernel.\newline $Y$ is the set of positions of all zeros in the kernel.}
  \Output{$C$ is the coefficient of incidence of the kernel.}
 $S \longleftarrow X$\;
 $r \longleftarrow 0$\;
 \For{$y\in Y\cap S^c $}{
$X' \longleftarrow X+y$\;
\tcp{\footnotesize{$X'$ is the set of positions of all ones in the kernel when the starting pixel is shifted to $y$}}
$r \longleftarrow r + |X'\cap S| + |X'\cap S^c\cap(X\cup Y)^c|$\;
$S \longleftarrow X'\cup S$\;
}
$C \longleftarrow \frac{r}{|X \cup Y|}$\;
\end{algorithm}

\subsection{Secret Share Generation}
In the proposed method, the kernel is just a selection mechanism that is being used to choose a permutation of image pixels along with a few random numbers. This is ensured by sequentially moving the kernel in the sequence of positions generated from the kernel itself. At each step, the kernel operation returns $k$ values (sequence of pixel data and random numbers) which are used as coefficients to generate the share polynomial in \eqref{eq1}. To avoid kernel from becoming SPOF for the entire scheme, it is shared along with the image across the $n$ participants. As the kernel data is uncorrelated it can be directly shared using Thien-Lin's \cite{Lin02} scheme. Kernel is also treated as a secret image and is divided into $n$ kernel shares using Thien-Lin's share generation procedure,
\begin{equation}
f_{kernel}(x) = (\gamma_0 + \gamma_1 x + \gamma_2 x^2 +\cdots+ \gamma_{k-1} x^{k-1}) \text{ mod} \ 2
\end{equation}
where the polynomial coefficients are the binary kernel data. Shares are generated from the polynomials and then the polynomials are destroyed. These shares are then distributed across the participants.

\subsection{Secret Image Reconstruction}
First step of reconstruction is to pool the kernel shares to regenerate the kernel using Thien-Lin's \cite{Lin02} secret reconstruction procedure. As the sequence of positions a kernel assumes while traversing the image is deterministic from the kernel itself, we can retrace the positions that the kernel takes. At every position of the kernel, $k$ or more image shares are collected to find the corresponding coefficients of the polynomial using Lagrange's interpolation. Then at every position of the kernel, the coefficients of the polynomial are mapped to the corresponding pixels, in the order in which they were selected by the kernel in the construction phase. Pixels that are already mapped must not be overwritten, as the later generated values correspond to the random coefficients. Same is the case for all the kernel indices that go out of image bounds, all these values correspond to random numbers which are to be discarded.



\section{Adversarial model}
The $(k,n)$ value, identities of the $n$ participants and the agent that is responsible for the share generation and distribution are assumed to be available to everyone in the model. It is assumed that the agent and the $n$ participants always follow the protocol. Also the communication between the agent and the participants during share distribution is assumed to be secure. Any communication that involves the agent requires authentication, thereby making it impossible for an adversary to mimic or replay an agent-participant interaction. This makes the agent a trusted entity among the participants. Most of these assumptions are valid in the case of data centers where the communications between the servers inside the facility are assumed to be secure. As the share distribution process is assumed to be secure, the adversary cannot possess a valid share. However, the adversary is capable of deceiving other participants to be one of the $n$ participants. So in this case, the adversary along with $k-1$ participants shall first pool their kernel shares to generate the kernel. As the shares used by the adversary are not authentic, the kernel generated from polynomial interpolation is highly improbable to satisfy the $(k,n)$ property (i.e sparse matrix with $k$ ones). This alerts the $k$ members that one among them is an adversary, but does not reveal who. Although the transaction does not proceed any further, the adversary can prevent other transactions by intervening in them. This problem can be avoided by implementing a verification scheme \cite{Lin10,Harn14}, where each participant must verify their credentials before the start of any transaction.

Assuming that the adversary is aware of the kernel size and image size, the probability of a random guess by the adversary to actually lead to a successful kernel share is,
\begin{equation}\label{kshareProb}
Pr(\textrm{Generating the correct kernel} |k,S) = \frac{1}{2^\frac{S}{k}}
\end{equation}
where S is the size of the kernel (as the kernel shares are $S/k$ bits long, and only one combination leads to perfect reconstruction). It can be seen from the \eqref{kshareProb} that increasing the kernel dimensions for a given $k$ provides greater security against random guessing attacks. The computational complexity of the kernel operation can be made independent of kernel dimensions by storing it as a sparse matrix. However, greater $S$ requires greater kernel share sizes. Hence it is crucial to choose an appropriate value for $S$ that balances the trade-off between scheme secrecy and storage efficiency. Further, in the secret image reconstruction phase, every share polynomial equation can have $p$ values ($p$ is taken as 251 in Thien-Lin's \cite{Lin02} method) and there are a total of $M / k$ equations where $M$ is the share size ($M$ also remains unknown to the adversary as the amount of randomness is controlled solely by the kernel). Assuming $M$ is known to the adversary,
\begin{equation}
Pr(\textrm{Generating the correct secret image} |k,S) = \frac{1}{p^{M/k}} .
\end{equation}

\noindent


\section{Analysis of the scheme}
The proposed method uses a fixed-sized kernel for secret sharing images of different dimensions. Even kernels with large dimensions can be stored as a sparse matrix, making them more memory efficient than permutation ciphers. Use of the kernel to mix a variable number of random numbers with pixel data makes the scheme more secure against attacks that rely on high spatial correlation in images. It is also implausible to reconstruct the original secret image back from its shares, using an incorrect kernel while following the procedure in the proposed method. This is because there is no other way to differentiate the random numbers from pixel data or obtain the sequence of positions without the kernel. The kernel alone defines the sizes of shares and the amount of randomness present in the coefficients of the share polynomial. To avoid the kernel form becoming a SPOF for the entire scheme, it is also shared across the participants using Thien-Lin's \cite{Lin02} method.

Although using a permutation cipher also reduces the correlation as suggested in \cite{Lin02} it becomes a SPOF for the scheme. The proposed method suggests that the kernel operation is not just memory efficient but also avoids any form of SPOF by sharing the kernel across the participants. Further, Thien-Lin's method always generates the same shares for a given secret image. This makes Thien-Lin's method vulnerable as even if one participant was compromised, the adversary shall have access to the entire history of the participant's shares of the shared secret images and consequently predict with very high probability which secret images were shared more than once. Alternatively, the proposed method generates different shares each time for the same secret image. This variability in shares is attributed to the random numbers introduced by  the kernel.

Share generation in a ($k,n$)-threshold secret sharing scheme has a computational complexity of order $O(nk)$. For reconstruction of polynomial coefficients from $k$ shares using Lagrange's interpolation, each coefficient demands a complexity of order $O(k)$. Hence, overall computational complexity of finding all the coefficients is of the order $O(k^2)$. Only when $k$ or more participants combine their kernel's shares they can reconstruct the kernel which in turn will be used for image reconstruction. With $k-1$ or fewer shares, it is not possible to reconstruct the kernel and subsequently the secret image (any interpolation technique would fail when the data is uncorrelated). It is advisable to use a sparse matrix as the kernel, as it will significantly reduce the random guessing attacks on the kernel.

The coefficient of incidence ($C$) correlates well with the  ratio  of  random  to  pixel  data, making it a quantifiable measure to compare and differentiate between different kernels. Security and storage space are two contrasting factors that generally exhibit a trade-off. For better security from attacks that exploit the correlation of the pixels in the image, a kernel with high $C$ must be chosen. This however will result in larger share sizes which may not be desirable. On the other hand, a kernel with low $C$ will result in efficient share sizes but make the scheme more vulnerable to attacks. One possible solution for a kernel with low $C$ is to choose the positions of ones in the kernel in a cyclic group over addition. When we are shifting the kernel to a new position, essentially we are shifting it to a position which is not a part of the group (as a new position must be an unmarked pixel). As translation of a kernel is an additive operation on position, the new set of ones also forms a cyclic group but has no element in common with the previous group. This drastically reduces the number of random coefficients in the share polynomial \eqref{eq1}. Further, increase in kernel size $(S)$ makes the scheme more secure \eqref{kshareProb} while also increases the share sizes.

\section{Conclusion}
The proposed $(k,n)$ threshold scheme uses Shamir's ($k,n$)-threshold secret sharing technique and its modification by Thien-Lin for sharing images. Kuang-Shyr Wu's method is optional as it further avoids truncation distortion along with providing a stronger immunity against random guessing attacks. Size of shares and the level of security are the key aspects of secret sharing. This paper discusses a general method that has versatile applications due to the flexibility offered by the kernel over these two key aspects. Moreover, the kernel's storage requirements are negligible when compared to image size, yet the kernel operation can be performed over images of any dimension. Further, the scheme is free from any form of SPOF. Coefficient of incidence is defined to help the users select a kernel that offers the desired share size and share security. The proposed method ensures that the compromise of any $k-1$ or fewer shares will not leak any information about the secret image. For further details about the procedure and comparison among various other secret image sharing schemes, we refer readers to the supplementary material at \href{https://akella17.github.io/kernel_papers/Supplementary.pdf}{https://akella17.github.io/kernel\_papers/Supplementary.pdf}.

\end{document}